\pdfoutput=1

\documentclass{appolb}

\usepackage{graphicx}

\usepackage{amsmath}
\usepackage{type1cm}
\usepackage{xspace} 
\usepackage{color}

\usepackage{amsmath,amssymb,bbm,color}

\newcommand{\afb}{{A_{\rm FB}}}
\newcommand{\alfqed}{{\alpha_{{}_{\rm QED}}}}
\newcommand{\vmax}{{v_{\max}}}

\newcommand{\kkmc}{{\tt   KKMC}}
\newcommand{\kkfoam}{{\tt KKFoam}}
\newcommand{\foam}{{\tt FOAM}}

\newcommand{\Rmf}{\mathfrak{R}}
\newcommand{\Mmf}{\mathfrak{M}}

\newcommand{\veps}{{\varepsilon}}

\begin{document}

\title{
\vspace{-4.0cm}
\begin{flushright}
		{\small {\bf IFJPAN-IV-2017-26}}   \\
\end{flushright}
\vspace{0.5cm}
Interference effects in a very precise 
measurement of the muon charge asymmetry at FCC-ee%
  \thanks{Presented by S.~Jadach at the ``Matter To The Deepest''
  Conference, Podlesice, Poland, 3--8~September~2017.
This work is partly supported by
 the grant of Narodowe Centrum Nauki  2016/23/B/ST2/03927 and The 
Citadel Foundation.
          }
} 

\author{S.\ Jadach$^a$ and S.~Yost$^{b}$
\address{$^a$ Institute of Nuclear Physics Polish Academy of Sciences,\\
              ul.\ Radzikowskiego 152, 31-342 Krak\'ow, Poland}
\address{$^b$ The Citadel, The Military College of South Carolina, \\
	171 Moultrie St., Charleston, SC 29409, USA }
}

\maketitle

\begin{abstract}
At the future high luminosity electron-positron collider FCC-ee proposed
for CERN, the precise measurement of the charge asymmetry
in the process $e^-e^+\to \mu^-\mu^+$ near the $Z$ resonance is of 
special interest.
In particular, such a measurement at $M_Z\pm 3.5$ GeV may provide
a very precise measurement of the electromagnetic coupling at the scale
$\sim M_Z$, a fundamental constant of the Standard Model.
However, this charge asymmetry is plagued by a large trivial contribution
from the interference of photon emission from initial state electrons
and final state muons.
We address the question whether this interference can be reliably calculated
and subtracted with the help of a resummed QED calculation.

\end{abstract}



\section{Introduction}
The Future Circular Collider with $e^\pm$ beams (FCC-ee) considered at CERN
could produce more that $10^{12}$ $Z$ bosons per year,
a factor of $10^5$ more than LEP collider, opening completely
new avenues in testing Standard Model (SM) predictions, 
which may reveal signals of New Physics beyond the Standard Model.
The pure electromagnetic coupling constant $\alfqed(M_Z)$,
will be vitally important in searching for disagreements between 
the FCC-ee experimental data and SM predictions 
at a precision level at least a factor of 10 better than in the past.
This kind of the measurement was proposed and analyzed
in ref.~\cite{Janot:2015gjr}.

Generally, $M_Z, G_F$, and $\alfqed(0)$
outweigh other data in the ``testing power''
in the overall fit of the SM to experimental data.
Up to now, $\alfqed(Q^2=0)$ was ported to $\alfqed(Q^2=M^2_{Z})$
using low-energy QCD data --
this limits its usefulness beyond LEP precision.
In ref.~\cite{Janot:2015gjr}, it was proposed
to use another observable, $A_{FB}(e^+e^-\to \mu^+\mu^-)$
at $\sqrt{s_\pm}=M_Z \pm 3.5$ GeV, because
it features a similar ''testing profile'' in the SM overall 
fit as  $\alfqed(M^2_{Z})$,
but could be measured very precisely at a high luminosity FCC-ee.%
\footnote{
  It is advertised as
  ``determining  $\alfqed(M^2_{Z})$ from $\afb(\sqrt{s_{\pm}})$''.
}
However, $\afb$ varies very strongly near $\sqrt{s_{\pm}}$, and
hence is prone to large QED corrections (for instance ISR).
In particular, away from the $Z$ peak, $\afb$ gets
a direct sizable contribution from 
QED initial-final state interference (IFI).
It is therefore necessary to re-discuss how efficiently
these trivial but large QED effects in $\afb$
can be controlled and/or eliminated.

In this context, the aim is to reduce QED uncertainty to
   $\delta \afb(e^+e^-\to \mu^+\mu^-) < 4\times 10^{-5}$.
Presently $\Delta\alfqed(M_Z)/\alfqed \simeq 1.1 \times 10^{-4}$,
using low-energy $e^+e^-$ data.
Recent studies using the same method of dispersion relations 
are quoting possible improvements down to
$\Delta \alpha/\alpha \simeq (0.5-0.2) \times 10^{-4}$.
To be competitive, $\afb$ has to provide $\Delta \alpha/\alpha < 10^{-4}$.
Using Fig.\ 4 of ref.~\cite{Janot:2015gjr},
$\Delta \alpha/\alpha < 10^{-4}$ 
translates into $\Delta\afb < 4 \times 10^{-5}$.
The LEP-era estimate of the QED uncertainty in $\afb$ outside the $Z$ peak
was $\sim 2.5 \times  10^{-3}$; see ref.~\cite{Z-physics-at-lep-1:89}.
Its improvement by a factor of 200 or more
sounds like a very ambitious goal!
However, there was an encouraging precedent: 
for QED photonic corrections to the $Z$ lineshape $(\sim 30\%)$,
the uncertainty was reduced down to 
$\delta \sigma /\sigma \simeq 3 \times 10^{-4}$;
see ref.~\cite{Jadach:1992aa}.

The general features of QED (photonic) corrections
in $\afb(e^+e^-\to \mu^+\mu^-)$ are the following.
Pure ISR (initial state radiation) has an indirect influence 
due to reduction of $\sqrt{s}$.
Non-soft higher order missing corrections are under very good control.
Pure FSR (final state radiation) is generally small for a sufficiently 
inclusive event selection (cut-offs),
but cut-off dependence has to be controlled with a high quality MC.
The direct contribution of IFI (initial-final state interference)
is suppressed at the peak but sizable off-peak.
The IFI effect comes from a non-trivial matrix-element, 
even in the soft-photon approximation.
The \kkmc\ Monte-Carlo program of refs.~\cite{Jadach:2000ir,Jadach:1999vf}
is the most sophisticated tool to calculate all the above effects.

\begin{figure}[!ht]
\centering
\includegraphics[width=60mm]{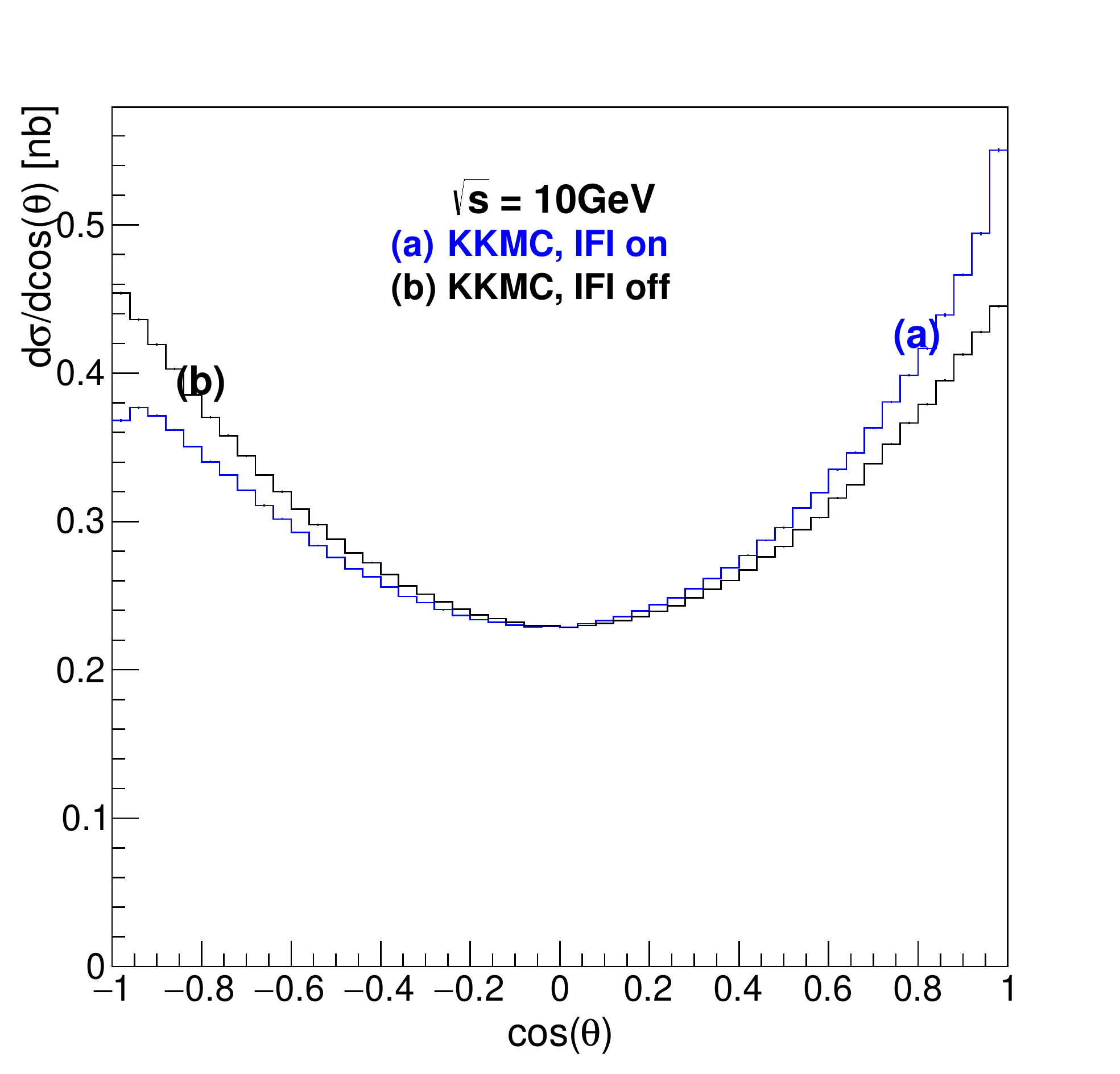}
\caption{\sf 
 Angular distribution of $e^-e^+\to \mu^-\mu^+$ at $\sqrt{s}=10$ GeV
 for (a) IFI switched on and (b) switched off.
}
  \label{fig:FIG1}
\end{figure}

\begin{figure}[!ht]
\centering
\includegraphics[width=120mm]{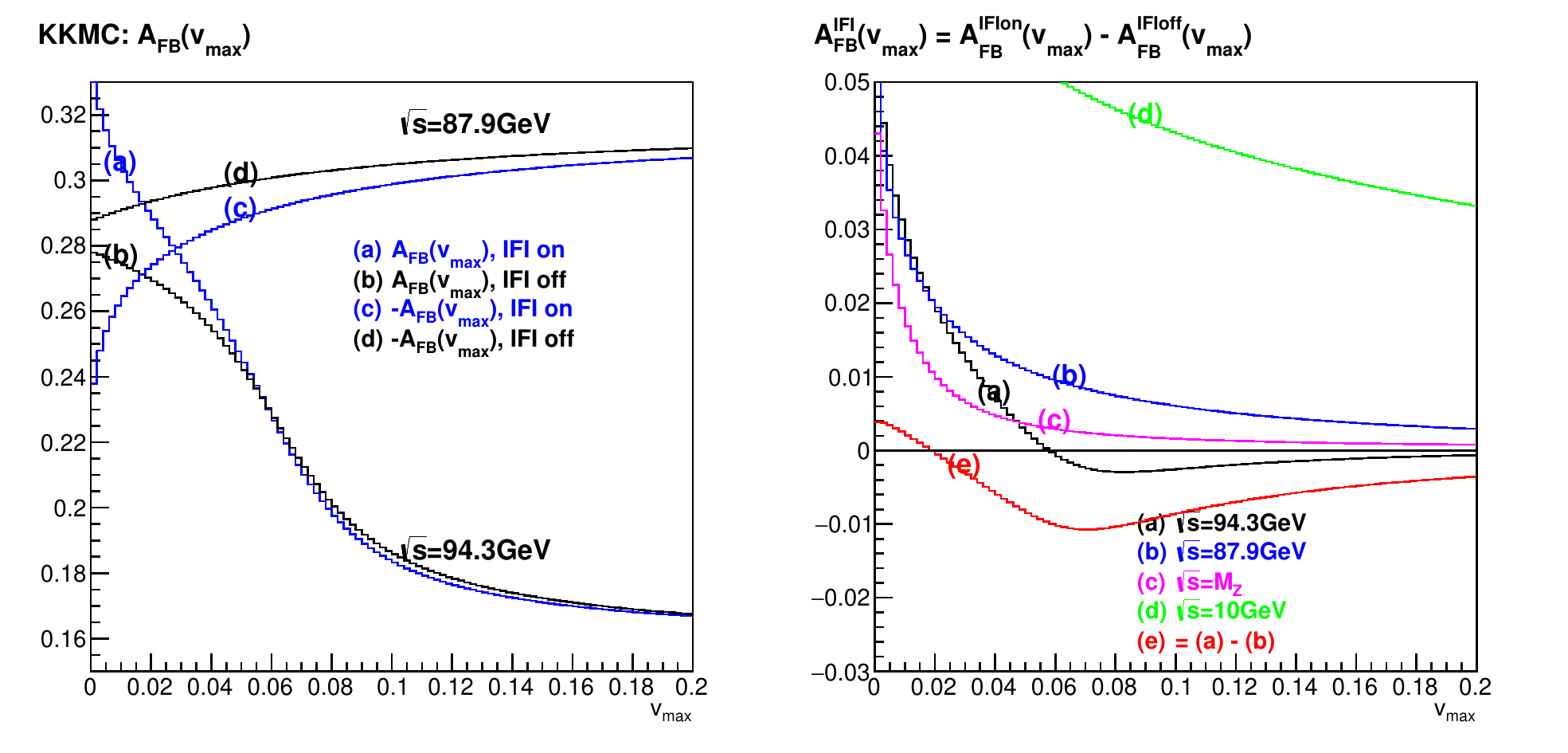}
\caption{\sf 
 Charge asymmetry in $e^-e^+\to \mu^-\mu^+$ at various energies
 as a function of the cut-off on $v=1-M^2_{\mu^+\mu^-}/s$,
 obtained using \kkmc.
}
  \label{fig:FIG2}
\end{figure}

A general understanding of the genuine IFI effect is the following:
In $e^-e^+\to \mu^-\mu^+$, not only is the $e^-$ annihilated, 
but its accompanying electromagnetic field of charge $-1$
also gets annihilated, and a new electromagnetic field of charge 
$-1$ is recreated along with the $\mu^-$.
At a wide scattering angle $\theta$, these two processes 
are independent sources of real photons.
The effect of the cut-off on the photon energy 
is therefore essentially $\theta$-independent.
For very small scattering angles, when the $\mu^-$ is close to the initial 
$e^-$, the $\mu^-$ is inherits most of its electromagnetic field from the 
$e^-$; hence real bremsstrahlung is weaker.
For $\theta\to 0$, the effect of the cut-off on the real photons
essentially vanishes.
On the other hand, in backward scattering ($\theta\to\pi$)
replacing the $e^-$ field of charge $-1$ with a $\mu^+$ field of charge $+1$ 
is ``more violent'', and more real photons are produced.
The effect of the cut-off on photon energy is then stronger and 
for sharper cut-off one gets
a pronounced dip in the muon angular distribution.

\section{IFI prediction from \kkmc}

The IFI phenomenon described above is clearly seen in the angular distribution
shown in Fig.~\ref{fig:FIG1}
for  $e^-e^+\to \mu^-\mu^+$ scattering at $\sqrt{s}=10$ GeV,
obtained using the \kkmc\ program.
Far from the resonance, the IFI contribution to charge asymmetry $\afb$
is about 3\% for a loose cut-off on photon energy $v<v_{\max}$,
and grows for stronger cut-offs; 
see line (d) in Fig.~\ref{fig:FIG2}.
Line (c) in Fig.~\ref{fig:FIG2} represents $\afb$ at $\sqrt{s}=M_Z$.
It illustrates the suppression of $\afb$ by the factor $\Gamma_Z/M_Z$
at the resonance position due to the time separation between
production and decay for any long-lived (narrow) resonance.
On the other hand, the two curves (a) and (b) in the RHS plot in 
Fig.~\ref{fig:FIG2} show the IFI contribution to
$\afb(v_{\max})$ at $\sqrt{s}=94.3$ GeV and 
$-\afb(v_{\max})$ at $\sqrt{s}=87.9$ GeV.
As we see, partial $\Gamma_Z/M_Z$ suppression is still in action.
The difference $(a)-(b)$ is also shown, demonstrating a partial cancellation
of IFI contributions between those two energies.

The \kkmc\ program of refs.~\cite{Jadach:2000ir,Jadach:1999vf}
provides the best state-of-the-art calculation of the IFI contribution
available today.
It includes an ${\cal O}(\alpha^2)$ QED photonic matrix element,%
\footnote{
  Except for the non-logarithmic parts of the ${\cal O}(\alpha^2)$
  IFI penta-boxes.
}
${\cal O}(\alpha^1)$ electroweak corrections,
soft photon resummation at the amplitude level,%
\footnote{
   A generalization of the Yennie-Frautschi-Suura exponentiation
   of ref.~\cite{yfs:1961}.
}
amplitude-level resummation of the $\sim\ln^n(\Gamma_Z/M_Z)$ QED effects due 
to the $Z$ resonance.%
\footnote{
  A generalization of the soft photon resummation near a resonance
  pioneered by the Frascati group; 
  see refs.~\cite{Greco:1975rm,Greco:1975ke,Greco:1980mh}.
}
Moreover, since \kkmc\ is regular Monte Carlo event generator,
it provides predictions for arbitrary experimental event selections,
cut-offs, and observables.

The ``problem'' with \kkmc\ predictions for IFI (and other observables)
is that there is no other calculation of comparable quality in order
to check for missing higher-order contributions and/or technical biases 
at the very high experimental precision anticipated at FCC-ee.

\section{New program \kkfoam\ for testing/calibrating \kkmc}

In order to address the above problem of calibrating and testing the
\kkmc\ predictions for IFI in $\afb$, 
another new program \kkfoam\ was recently developed.
It is based on soft-photon resummation including resonance effects
as in refs.~\cite{Greco:1975rm,Greco:1975ke,Greco:1980mh},
integrating analytically over photon angles,
and using the \foam\ Monte Carlo simulator/integrator \cite{foam:2002},
to integrate the remaining four photon energy variables
and muon scattering angle.
\kkfoam\ can calculate distributions only for a very limited
class of experimental cuts, on $v$ and $\cos\theta$.

\begin{figure}[!t]
  \centering
  \includegraphics[width=0.80\textwidth]{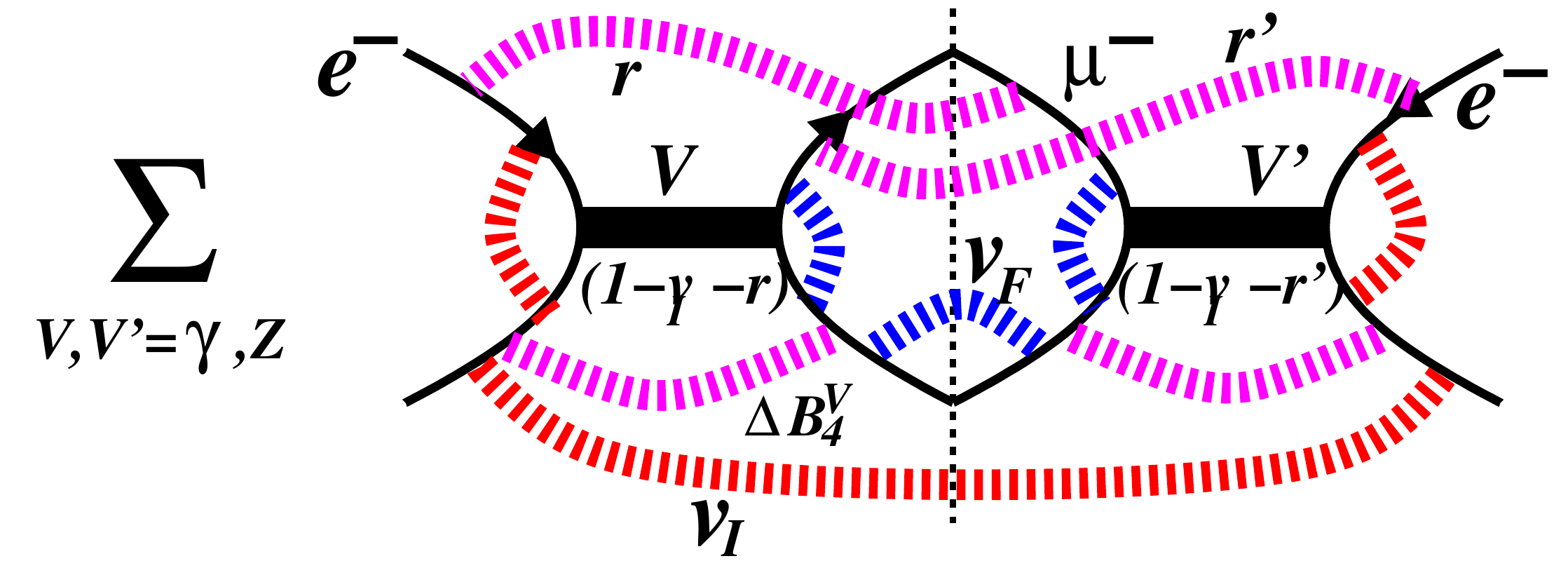}
  \caption{\sf
  Exponentiated multiple photon emission from initial and final fermions
  including ISR, FSR and IFI in the resonant process, 
  as in eq.~(\ref{eq:compact}).
  Dashed lines represent multiple real and/or virtual photons.
  }
  \label{fig:xKKsemIFI}
\end{figure}

A detailed description of the QED distributions used in \kkfoam\
will be available in a forthcoming publication~\cite{ifjpan-2017-11}
-- here we will only describe briefly its soft photon content
in a simplified form. It reads as follows:
\begin{equation}
\begin{split}
&\sigma(s,v_{\max})=
\frac{3\sigma_0(s)}{16}
\sum_{V,V'}
\int_0^1 dv_I\;  dv_F\; dr\; dr'\;
\\&~
\int d \cos\theta d\phi\; \theta(v_{\max}-v_I-r-r'-v_F)
\\&~~~
 \rho(\gamma_I,v_I)\; \rho(\gamma_X,r)\;\rho(\gamma_X,r')\; \rho(\gamma_F,v_F)\;
 e^{Y(p_i,q_i)}\;
\\&~~~
 \frac{1}{4}\sum_{\veps\tau}
 \Rmf\big\{e^{\alpha \Delta B_4^V(s(1-v_I-r))}   
           \Mmf^V_{\veps\tau}\big(v_I+r,\cos\theta\big)\;
\\&~~~~~~~~~~~~~~~~~~~~~~~~~~\times
          [e^{\alpha \Delta B_4^{V'}(s(1-v_I-r'))}
           \Mmf^{V'}_{\veps\tau}\big(v_I+r',\cos\theta\big)]^*
    \big\},
\label{eq:compact}
\end{split}
\end{equation}
with
\begin{equation}
\begin{split}
&\rho(\gamma,v) \equiv F(\gamma)\gamma v^{\gamma-1} 
	=  \frac{v^{\gamma-1} e^{-\gamma C_E}}{\Gamma(\gamma)},
\\&
\gamma_I(s) = \int\frac{d^3 k}{k^0} S_I(k)
	\delta\left(\frac{2k^0}{\sqrt{s}}-1\right), 
\\&
\gamma_F(s(1-v_I)) = \int\frac{d^3 k}{k^0} S_F(k)
	\delta\left(\frac{2k^0}{\sqrt{s}}-1\right), 
\\&
\gamma_X(\cos\theta) = \int\frac{d^3 k}{k^0} S_X(k)
	\delta\left(\frac{2k^0}{\sqrt{s}}-1\right), 
\end{split}
\end{equation}
and the classic YFS form factor
\begin{equation}
Y(s,t) = 2\alpha\Rmf B_4(s,t,m_\gamma) + \int_{2k^0<\sqrt{s}} 
\frac{d^3 k}{k^0}\left[S_I(k) + 2S_X(k) + S_F(k)\right],
\end{equation}
which is finite in the $m_\gamma\rightarrow 0$ limit.
Here, $\theta$ is angle the between the momenta of the $e^-$ and $\mu^-$,
$\sigma_0$ is the point-like cross section, 
$S_I, S_F, S_X$ are the usual eikonal factors \cite{Jadach:2000ir}
for photon emission from the initial state, final state, 
and their interference, 
$B_4$ is standard virtual Yennie-Frautschi-Suura formfactor \cite{yfs:1961}
and $\Mmf^{V}_{\veps\tau},\; \veps,\tau=\pm 1$ are Born spin amplitudes.
The additional form-factor $\Delta B_4^Z(s)$ due to the $Z$ resonance
is that of eq.~(94) in ref.~\cite{Jadach:2000ir},
while $\Delta B_4^\gamma(s)=0$.
The structure of the above distribution is illustrated 
schematically in Fig.~\ref{fig:xKKsemIFI}.
The remarkable feature is that there are independent photon energy variables 
$r$ and $r'$ for the matrix element $\Mmf$ and its complex conjugate!
A more detailed description of the QED distributions in \kkfoam\
will be provided in ref.~\cite{ifjpan-2017-11}.

\begin{figure}[!ht]
\centering
\includegraphics[width=120mm]{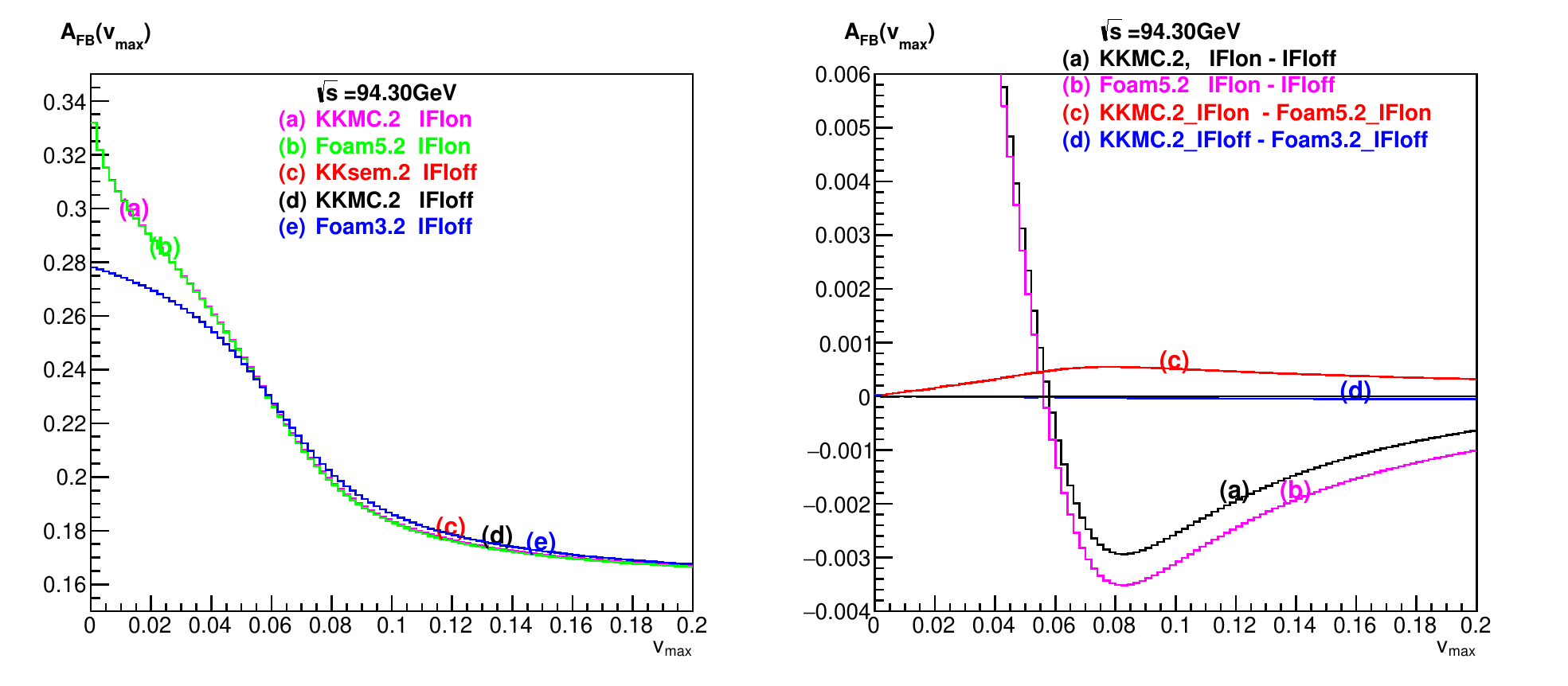}
\caption{\sf 
 Muon charge asymmetry at $\sqrt{s}=94.3$ GeV
 as a function of the cut-off $v_{\max}$ on the total photon energy.
 Results are obtained using \kkmc\ and \kkfoam.
}
  \label{fig:FIG3}
\end{figure}

\section{Testing IFI from \kkmc\ using \kkfoam}

With \kkfoam\ in hand, we can compare the predictions for the IFI
contribution to $\afb$ to those from \kkmc.
An example of such a comparison is shown in Fig.~\ref{fig:FIG3}, 
where the muon charge asymmetry is displayed
as a function of the total photon energy cut-off $\vmax$.
With IFI on, the difference (line (c) in the RHS plot)
is of order $4\cdot 10^{-4}$ and vanishes
as it should at $\vmax\to 0$.
This difference is a factor of 10 smaller than any QED uncertainty
due to IFI quoted in the LEP era, and it represents definite
progress.  However, it is still far (by another factor 10) from what
is needed for the FCC-ee experiments.
Nevertheless, a new avenue is now opened towards this ambitious goal.

\section{Summary}

We have presented an example of a promising new 
calculation of the initial-final QED interference
contribution to charge asymmetry in the $e^-e^+$ $\to$ $\mu^-\mu^+$
process near the $Z$ resonance, which brings us closer 
to mastering this effect 
at the precision level needed in the FCC-ee project --
that is, a factor of 10 better than the LEP-era state of the art.


\providecommand{\href}[2]{#2}


\end{document}